\documentclass[aps,prmaterials,reprint,amsmath,amssymb]{revtex4-2}

\usepackage[utf8]{inputenc}
\usepackage{graphicx}
\usepackage{subcaption}
\usepackage{setspace}
\usepackage{placeins}
\usepackage{booktabs}
\usepackage{hyperref}
\hypersetup{colorlinks,allcolors=black}
\usepackage{nameref}
\usepackage{lineno}

\DeclareUnicodeCharacter{2212}{-}

\graphicspath{{figures/}}

\begin{document}
\doublespacing

\title{Cooperative \texorpdfstring{CO$_2$}{} capture via oxalate formation on metal-decorated graphene}

\date{\today}
\author{Inioluwa C. Popoola}
\affiliation{Yusuf Hamied Department of Chemistry, University of Cambridge, United Kingdom}

\author{Benjamin X. Shi}
\affiliation{Yusuf Hamied Department of Chemistry, University of Cambridge, United Kingdom}

\author{Fabian Berger}
\affiliation{Yusuf Hamied Department of Chemistry, University of Cambridge, United Kingdom}

\author{Andrea Zen}
\affiliation{Dipartimento di Fisica Ettore Pancini, Universita di Napoli Federico II, Monte Sant’Angelo, I-80126 Napoli, Italy }
\affiliation{Department of Earth Sciences, University College London, London WC1E 6BT, United Kingdom}

\author{Dario Alfè}

\affiliation{Dipartimento di Fisica Ettore Pancini, Universita di Napoli Federico II, Monte Sant’Angelo, I-80126 Napoli, Italy }
\affiliation{London Centre for Nanotechnology, University College London, London WC1E 6BT,
United Kingdom}
\affiliation{Thomas Young Centre, University College London, London WC1E 6BT, United Kingdom }
\affiliation{Department of Earth Sciences, University College London, London WC1E 6BT, United Kingdom}

\author{Angelos Michaelides
}
\affiliation{Yusuf Hamied Department of Chemistry, University of Cambridge, United Kingdom}

\author{Yasmine S. Al-Hamdani}

\affiliation{Thomas Young Centre, University College London, London WC1E 6BT, United Kingdom }
\affiliation{Department of Earth Sciences, University College London, London WC1E 6BT, United Kingdom}

\begin{abstract}
    % Your abstract goes here
   CO$_2$ capture using carbon-based materials, particularly graphene and graphene-like materials, is a promising strategy to deal with CO$_2$ emissions. However, significant gaps remain in our understanding of the molecular-level interaction between CO$_2$ molecules and graphene, particularly, in terms of chemical bonding and electron transfer. In this work, we employ random structure search and density functional theory to understand the adsorption of CO$_2$ molecules on Ca, Sr, Na, K, and Ti decorated graphene surfaces. Compared to the pristine material, we observe enhanced CO$_2$ adsorption on the decorated graphene surfaces. Particularly on group 2 metals and titanium decorated graphene, CO$_2$ can be strongly chemisorbed as a bent CO$_2$ anion or as an oxalate, depending on the number of CO$_2$ molecules. Electronic structure analysis reveals the adsorption mechanism to involve an ionic charge transfer from the metal adatom to the adsorbed CO$_2$. Overall, this study suggests that reducing CO$_2$ to oxalate on group 2 metals and titanium metal-decorated graphene surfaces is a potential strategy for CO$_2$ storage.
\end{abstract}

\maketitle
\onecolumngrid
%\begin{abstract}
    
%\end{abstract}

\section{Introduction}

%Understanding the adsorption chemistry of CO$_2$ on surfaces is of important because of the significant contribution of CO$_2$ to climate change and
The world faces a severe climate crisis due to the alarming increase of anthropogenic CO$_2$ emissions. One key strategy to offset CO$_2$ emissions is to identify new and high-performing materials capable of efficiently trapping CO$_2$ molecules within an optimal adsorption energy range of $-0.4$ to $-0.8$ eV \cite{belmabkhout2016low, Wang2021, Lu2022, Darvishnejad2020}. This range is crucial for ensuring moderate binding and subsequent release of CO$_2$, facilitating its conversion into valuable chemicals \cite{belmabkhout2016low, Wang2021, Lu2022, Darvishnejad2020}.

A wide range of porous materials, such as activated carbon, zeolites, graphene, and metal-organic frameworks have been investigated as promising candidates for CO$_2$ capture because of their high surface areas,  stability, and tunable surface chemistries \cite{kim2020cooperative,lu2008comparative,rashidi2016overview, balasubramanian2015recent,malko2012competition,ferrari2006raman}.  
In particular, graphene is a two-dimensional material with great promise for many applications such as nanoelectronics \cite{park2016graphene, xuan2008atomic}, gas storage \cite{gadipelli2015graphene,szczkesniak2017gas, al2023mechanisms}, chemical sensors \cite{liu2012biological,fowler2009practical} and catalytic applications \cite{allen2010honeycomb, gilje2007chemical}. The versatility of graphene can be attributed to its unique electronic structure. Specifically, the semi-metallic nature of graphene as a result of its vanishing density of states at the Fermi level enables the electronic properties of this material to be easily tuned \cite{geim2007rise, geim2009graphene,kattel2012stability}.

As promising as graphene is for gas storage applications, CO$_2$ has been reported to bind too weakly on pristine graphene below the lower adsorption energy limit for viable CO$_2$ capture \cite{Wang2021, Lu2022, Darvishnejad2020}. Hence, there is a need to functionalise the surface of graphene.
Owing to the reducibility of CO$_2$ and the tunability of graphene's electronic properties, decoration of graphene with neutral metal adatoms has been found to enhance the interaction between CO$_2$ and graphene \cite{lu2014adsorption, Cazorla2011}. However, this enhanced interaction is typically seen when one CO$_2$ molecule is adsorbed and little is known about the adsorption of multiple CO$_2$ molecules on the metal-decorated graphene surface \cite{Cazorla2011, liu2019synergetic, Vallejo2022, promthong2020transition, cortes2018fe, ma2019doping, Wang2021, Lu2022, tawfik2015multiple}. 
Investigating the interaction of multiple CO$_2$ on decorated graphene surfaces could help in understanding the cooperative interaction between CO$_2$ molecules and provide fundamental insights that aid in the discovery of next-generation materials for an enhanced CO$_2$ uptake.

In this work, we investigate the adsorption of multiple CO$_2$ molecules on metal-decorated graphene (M@Gr) surfaces, identify high-performing metal decorators and elucidate the underlying mechanism mediating the interaction between the gas molecules and surfaces. To accurately predict the M@Gr properties, we used density functional theory (DFT) and an \textit{ab initio} random structure search (RSS) approach \cite{pickard2011ab,oganov2019structure, mcmahon2011ground} to systematically find the most stable configuration of 1-7 CO$_2$ molecules adsorbed on group 1 metals, Na and K, group 2 metals, Ca and Sr, and transition metal, Ti, decorated graphene surfaces. 

The results computed at the PBE-D3 \cite{moellmann2014dft, perdew1996generalized} level of theory show that decorating graphene with a metal atom can boost the adsorption strength to $-$1.80 eV -- about ten times greater than the pristine material ($-$0.17 eV). %, exceeding the upper adsorption energy limit of $-$0.8 eV for viable CO$_2$ capture.
We find that the enhanced interaction for group 2 and Ti decorators results from the one-electron reduction of CO$_2$ to a bent CO$_2$ radical anion and the formation of oxalate when $\geq$ 2 CO$_2$  molecules are adsorbed.  The oxalate further enhances the cooperative adsorption of subsequent physisorbed CO$_2$ molecules due to a strong interaction of CO$_2$ with the newly formed metal cation in the system. This highlights the propensity of group 2 and Ti-decorated graphene for the adsorption of multiple CO$_2$ molecules.

The remainder of this paper is organized as follows: We provide the computational details of the DFT and the RSS approach in Section II and discuss the adsorption energy trends,  formation of oxalate and the underlying adsorption mechanism in Section III. In Section IV, we briefly discuss our results in context with previous works, and in Section V, we end with conclusions and an outlook on future related research.
%suggest future direction for research focusing on the practical application of M@Gr for CO$_2$ capture

\section{Computational details}\label{sec:methods}

 All calculations were performed with the plane-wave DFT code, Vienna \textit{ab initio} simulation package (VASP5, version 5.4) \cite{hafner2008ab, kresse1993ab,kresse1996efficient, kresse1994ab}. The Perdew, Burke, and Ernzerhof (PBE) exchange-correlation functional \cite{perdew1996generalized}, and Grimme's D3 dispersion correction with zero damping function \cite{Grimme2010} were used. This combination was chosen because PBE-D3 can successfully describe interactions between gas molecules and surfaces \cite{araujo2022adsorption, ma2011adsorption,brandenburg2019interaction,al2017properties}. In addition, the adsorption trends reported in this study are found to be rather insensitive to functional choice -- see the Supporting Information (SI) for details. Projector augmented wave (PAW) potentials were used to describe the valence-core interaction \cite{blochl1994projector,kresse1999ultrasoft}. The 2s2p, 2s2p, 3s3p4s (Ca$_{sv}$), 4s4p5s (Sr$_{sv}$),  2s2p3s (Na$_{sv}$), 3s3p4s (K$_{sv}$) and 3s3p4s3d (Ti$_{sv}$) electrons were explicitly included for C, O, Ca, Sr, Na, K, and Ti atoms, respectively.
%The C, O, Ca$_{sv}$, Sr$_{sv}$, Na$_{sv}$, K$_{sv}$, and Ti$_{sv}$ pseudopotentials are used in this work.

A 5 $\times$ 5  supercell of graphene containing 50 carbon atoms was used, except where otherwise stated. Periodic boundary conditions were applied, and a vacuum of 20 Å was used in the $z$-direction, perpendicular to the graphene sheet. We used a low metal adatom doping concentration of 2\% in this study. This concentration refers to the ratio between the number of metal adatoms and carbon atoms in graphene. Specifically, 2\% translates to one metal adatom per 5 $\times$ 5 graphene supercell (50 C atoms). 
We have selected this low concentration to obtain insights into the individual CO$_2$ adsorption processes on decorated graphene.

The lowest energy structures are the most important for predicting properties of materials because they tend to dominate under equilibrium conditions.
To identify the most stable conformers for the adsorption of 1-7 CO$_2$ molecules on the Ca, Sr, K, Na, and Ti decorated graphene (from here on referred to as Ca@Gr, Sr@Gr, K@Gr, Na@Gr and Ti@Gr, respectively), we screened the adsorption energy of CO$_2$ on each substrate in a two-stage process.
In the initial screening stage, 100 random configurations each for 1-7 CO$_2$ molecules were generated on Ca, Sr, K, Na, and Ti@Gr systems.  The atoms in the graphene sheets were fixed, and the RSS was done with both linear and bent CO$_2$ molecules, which were placed at a random distance and angle on the $xy$-plane of the graphene sheet. The distance, $r$, between the metal adatom and the carbon atom in CO$_2$ in the z-direction was constrained to be within 2 $\leq$ $r$ $\leq$ 4 Å. A spin-polarized $\Gamma$ point relaxation was performed for all systems using a plane-wave energy cut-off of 300 eV and a Gaussian smearing of 0.05 eV. All geometry and electronic relaxations for 1-7 CO$_2$ molecules on each substrate were converged to residual forces and energy less than  0.05 eV/Å and 1 $\times$ 10$^{-5}$ eV, respectively. %This screening gives a range of high, moderate and low energy structures. 

In the refined screening, an energy cut-off of 400 eV and a \textbf{k}-point mesh of 3 $\times$ 3 $\times$ 1 were used to obtain more accurate adsorption energies for the lowest energy structures from the initial screening stage. At this setting, the adsorption energy is converged within 10 and 1 meV for the \textbf{k}-point and the energy cut-off, respectively (see Fig. S6 in SI for \textbf{k}-point and energy cut-off convergence). All geometry and electronic relaxations were converged to residual forces and energy less than 0.01 eV/Å and 1 $\times$ 10$^{-6}$ eV, respectively.

The adsorption energies were computed using Eqn. \ref{Eqn_1} and reported in Table. \ref{Table_Binding_energy}.

\begin{equation}\label{Eqn_1}
     E_{ads} = \frac{E^{tot}_{M@Gr+\textit{n}CO_{2}} - E^{tot}_{M@Gr} - \textit{n}E^{tot}_{CO_{2}}}{\textit{n}}
\end{equation}

\noindent In Eqn. \ref{Eqn_1}, $E_{ads}$ is the average adsorption energy per CO$_2$ molecule, ${E^{tot}_{M@Gr+nCO_{2}}}$ is the total energy of CO$_2$ molecules adsorbed on M@Gr, $E^{tot}_{M@Gr}$  is the total energy of the fully relaxed substrate, $E^{tot}_{CO_{2}}$ is the total energy of the gas-phase relaxed CO$_2$ molecule and ${n}$ is the number of CO$_2$ molecules adsorbed. 

To understand the charge transfer processes for CO$_2$ adsorption on different metal atoms, a charge density distribution analysis was done on the lowest energy structures using the Bader charge approach \cite{bader1985atoms}. %The density of state analysis (DOS)/ projected density of state (PDOS) 
Density of states (DOS) and projected density of states (PDOS) were also examined and these were obtained using a $\Gamma$-centered 15 $\times$ 15 $\times $1 \textbf{k}-point mesh and the SUMO code for post-processing the data \cite{ganose2018sumo}.

\section{Results}\label{sec:Results}

\begin{figure}
    \centering
    \includegraphics[width=1\textwidth]{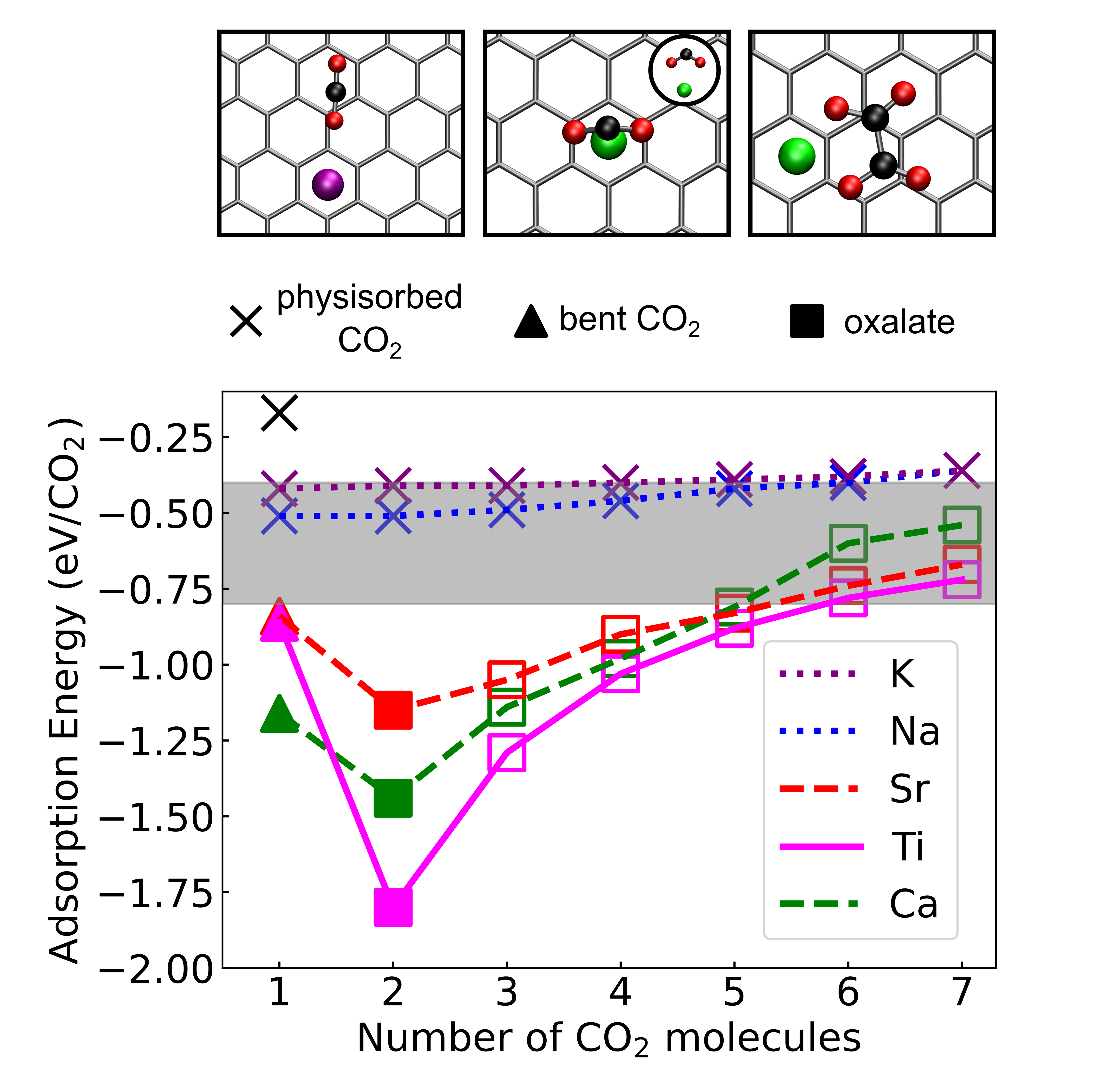}
    \captionsetup{justification=raggedright,singlelinecheck=false}
    \caption{Adsorption energy plot of 1-7 CO$_2$ molecules adsorbed on metal decorated graphene.
    The average adsorption energies shown herein (eV) are the converged energy for the lowest energy structures found with the RSS. The symbol indicates the optimized orientation of CO$_2$ molecules around the metal atoms. The triangles indicate the formation of a bent CO$_2$ anion; the crosses indicate the radial orientation of the physisorbed CO$_2$ molecules to the metal atom. The filled squares indicate the exclusive formation of oxalate, and the unfilled squares represent the system with oxalate and physisorbed CO$_2$ molecules.  The single black cross represents the adsorption energy of CO$_2$ on pristine graphene. Examples of optimized orientations of CO$_2$ on the substrate are shown above the plot. The dotted lines represent group 1 elements; the dashed lines indicate group 2 elements; and the solid line shows the transition metal, Ti. Different colours show different metal decorators. The shaded gray region shows the ideal adsorption and desorption energy window ($-$0.4 to $-$0.8 eV) for carbon capture \cite{Chu2012}. }
    \label{Binding_energy_plot}
\end{figure}

Our RSS  procedure was used to obtain the most stable configuration of 1-7 CO$_2$ molecules adsorbed on Ca, Sr, Na, K and Ti@Gr, and the adsorption energies computed for the most stable configurations of adsorbed CO$_2$ on each substrate are reported in Table \ref{Table_Binding_energy} and shown in Fig.~\ref{Binding_energy_plot}. The metal atoms are seen to be adsorbed at about 2.0 - 2.8 Å on the hollow site of the graphene sheet with adsorption energies reported in Table  \ref{Table_Binding_energy} \cite{Yasmine2023}.
% on graphene compared to the cohesive energies of the corresponding bulk metals\cite{jana2018assessing,Yasmine2023,cazorla2010first,Cazorla2011}. The cohesive energy of the bulk metal
In contrast to the weak interaction of CO$_2$ with pristine graphene ($-$0.17 eV), the interaction of CO$_2$ with M@Gr systems is generally much stronger for all metal decorators. %than the ideal adsorption energy window of $-$0.4 to $-$0.8 eV (shown as the grew region in Fig.~\ref{Binding_energy_plot}).
The average adsorption energies range from $-$0.54 to $-$1.44 eV for group 2 decorated graphene, $-$0.36 to $-$0.51 eV for group 1 decorated graphene, and $-$0.72 to $-$1.80 eV for Ti decorated graphene. The strongest adsorption of CO$_2$ on group 2 and Ti metal decorators is observed when two CO$_2$ molecules are adsorbed, beyond which the average adsorption energy decreases as subsequent CO$_2$ molecules are physisorbed, while the total adsorption increases (see Fig. S2 in SI). The adsorption strength of the metal decorators follows the order of Ti $>$ Ca $>$ Sr $>$ Na $>$ K, with Ti exhibiting the strongest adsorption and K the weakest.  Overall, the adsorption energy of group 2 decorated graphene and the Ti@Gr is outside the ideal adsorption energy window  ($-$0.4 to $-$0.8 eV) for adsorption and desorption of CO$_2$ (see the shaded grey region in Fig. \ref{Binding_energy_plot}), while the adsorption on group 1 decorated graphene falls within the range.

\subsection{Enhanced \texorpdfstring{CO$_2$}{} adsorption via metal-decoration of graphene}\label{sec:multiple CO2 on M@gr}

CO$_2$ has been previously reported to be weakly physisorbed on pristine graphene, exhibiting an adsorption energy of about $-$0.2 eV, which is below the adsorption limit for viable CO$_2$ capture \cite{Wang2021, Lu2022, Darvishnejad2020}. The most stable configuration of CO$_2$ on pristine graphene from our RSS result is a linear CO$_2$ molecule, flatly adsorbed on a C-C bond of pristine graphene with a weak adsorption energy of $-$0.17 eV (see Fig. \ref{Binding_energy_plot}) and at a distance of 3.4 Å  %to the nearest carbon on%
 to the graphene sheet (see the inset in Fig. S2 of the SI). The equilibrium C-O bond length of 1.18 Å remains unchanged upon adsorption, indicating a weak interaction of CO$_2$ on the graphene sheet. This result agrees with previous studies, which reported weak physisorption of CO$_2$ on pristine graphene  \cite{Wang2021, Lu2022, Darvishnejad2020}.

\begin{table*}
\centering
\captionsetup{justification=raggedright,singlelinecheck=false}
\caption[Adsorption energies of 1-7 CO$_2$ molecules adsorbed on group 1, group 2 and transition metal-decorated graphene.]{Adsorption energies of 1-7 CO$_2$ molecules adsorbed on group 1, group 2 and transition metal-decorated graphene. The adsorption energies  ($E_{M-Gr}^{ads}$) and metal-graphene distances ($d_{M-Gr}$) in the absence of CO$_2$ are also reported, with some numbers, labelled $^a$ , reported from Ref. \cite{Yasmine2023}}
\begin{minipage}{\textwidth}
  \centering
    \begin{tabular}{@{}llllllllll@{}}
    \toprule
Adatom   & 1     & 2    & 3     & 4     & 5    & 6    & 7   &   $d_{M-Gr}$ (Å)        & $E_{M-Gr}^{ads}$ (eV)     \\ \midrule
Ca      & -1.16 & -1.44 & -1.14 & -0.98 & -0.81 & -0.60 & -0.54  & 2.76\footnotemark[1] & −0.74\footnotemark[1]\\
Sr    & -0.84 & -1.15 & -1.05 & -0.90 & -0.83 & -0.74 & -0.67  & 2.50\footnotemark[1] & −0.75\footnotemark[1]\\
K    & -0.42 & -0.41 & -0.41 & -0.40 & -0.39 & -0.38 & -0.36 & 2.57\footnotemark[1] & −1.20\footnotemark[1]\\
Na  & -0.51 & -0.51 & -0.49 & -0.46 & -0.42 & -0.40 & -0.36  & 2.19\footnotemark[1] & −0.72\footnotemark[1]\\
Ti   & -0.86 & -1.80 & -1.29 & -1.03 & -0.88 & -0.78 & -0.72 & 2.35    & -1.94 \\ \bottomrule
    \end{tabular}
\end{minipage}
    \label{Table_Binding_energy}
\end{table*}

%\footnotetext{Reported from Ref. \cite{Yasmine2023}}

%\subsection{Adsorption of one CO$_2$ molecule on metal-decorated graphene}\label{1co2@M@Gr}($\theta$(O-C-O)) =  

For the first adsorbed CO$_2$ molecule on Ca and Sr@Gr, the CO$_2$ molecule undergoes a significant bending and symmetrical C-O bond elongation to  125\textdegree ~ and 1.28 Å for Ca. Similar bending of CO$_2$ with an angle of 134\textdegree~ and bond elongation to 1.30 Å is seen for Sr. This change in structure is as a result of charge transfer of $\approx$ 1 e$^-$ from the metal adatom to the pi orbital of the adsorbed CO$_2$ with no hybridization of the p-state and the s,d metallic states of the Ca adatom as reported in Ref \cite{Cazorla2011}, (see Fig. S5 (b) in SI). The rearrangement of the linear CO$_2$ to the bent CO$_2$ anion on group 2 decorated graphene and the observed structural changes in this study agree well with previous studies that investigated the adsorption of CO$_2$ on Ca@Gr \cite{Cazorla2011, tawfik2015multiple}. 

The bent CO$_2$ anion formation on the group 2 decorated graphene results in the chemisorption energies reported for 1 CO$_2$ on Ca and Sr in Table \ref{Table_Binding_energy}. Hoffmann \cite{hoffmann1988chemical} describes the weakening and elongation seen for the intramolecular C-O bond length as the chemisorption compromise, explaining the bonding between a surface and an adsorbate. The strengthening of the interaction between the adsorbate and the surface occurs at the expense of the bonding within the M@Gr and the adsorbed molecules (CO$_2$) \cite{hoffmann1988chemical}. This increases the intramolecular bond length within the adsorbates and surfaces involved, just as seen for the bent CO$_2$ in this work.

For the adsorption of one CO$_2$ on the Ti@Gr, the lowest energy structure is a partially dissociated CO and O molecule on the surface with a strong adsorption energy of -2.96 eV per CO$_2$ molecule. The dissociation of CO$_2$ on Ti@Gr occurs due to a charge transfer of more than one e$^-$ from the Ti adatom to the CO$_2$ molecule. This is typically an unwanted molecular dissociation that is well-known both experimentally and theoretically with Ti decorators, which limits their gas storage potential \cite{shevlin2008high, yildirim2005titanium, zhao2005hydrogen,cazorla2010first}.  The dissociative adsorption energy of a single CO$_2$ molecule on Ti@Gr was not reported in Table \ref{Table_Binding_energy} because it is about three times the upper limit of the ideal adsorption energy window for  CO$_2$ capture shown in Fig. \ref{Binding_energy_plot}. In addition, in the presence of two or more CO$_2$ molecules, the partially dissociated CO$_2$ molecule is not the most stable configuration. As such, we focus on the bent CO$_2$ anion within this energy window. The adsorption energy for the bent  CO$_2$ molecule is reported in Table \ref{Table_Binding_energy} and the bent CO$_2$ has a C-O bond length of 1.29~Å, an angle of  135\textdegree~and is adsorbed at a distance of 1.64 Å to the Ti adatom. 

For the adsorption of one CO$_2$ molecule on Na and K, there is no reduction of the linear CO$_2$ molecule to a bent CO$_2$ anion, despite these adatoms having the required one valence electron to reduce CO$_2$ to a bent CO$_2$ anion. Instead, graphene is reduced by the valence electrons of these metal atoms and the negatively charged graphene layer does not possess a sufficiently strong reduction potential to reduce the CO$_2$. The most stable configuration of CO$_2$ on group 1 decorated graphene is found to be radially coordinated to the adatom (see inset label physisorbed CO$_2$ in Fig. \ref{Binding_energy_plot}) with the nearest oxygen atom at a distance of 2.34~Å, with an unchanged C-O bond length of 1.18 Å. This structural reorientation of CO$_2$ on group 1 decorated graphene suggests an electrostatic interaction of the adsorbed CO$_2$ with the oxidized group 1 cations in the system. The lack of bent CO$_2$ anion formation on group 1 decorators explains the relatively weak physisorption energy seen for these systems compared to other metal decorators \cite{ma2019doping}.

\subsection{Enhanced interaction of \texorpdfstring{CO$_2$}{}  via oxalate formation}
\label{sec:oxalate formation}

 For the adsorption of two CO$_2$ molecules on group 2 decorated graphene, RSS found an oxalate with an average adsorption energy of $-$1.44, $-$1.15, and $-$1.80 eV for Ca@Gr,  Sr@Gr, and Ti@Gr, respectively. In the oxalate, 2 CO$_2$ molecules form a carbon-carbon covalent bond of 1.6  Å length. The oxalate coordinates with the metal ion and carries a charge of about $-$2. %Tawfik \textit{et al.} \cite{tawfik2015multiple}, have reported the adsorption of 1 - 5 CO$_2$ molecules on metal-doped graphene. The finding from Tawfik's \cite{tawfik2015multiple} for the adsorption of two CO$_2$ molecules on group 2  and TM@Gr is inconsistent with our findings here. In Ref.\cite{tawfik2015multiple}, only the first adsorbed CO$_2$ formed a bent CO$_2$ anion, while subsequently adsorbed CO$_2$ were physisorbed \cite{tawfik2015multiple}.  %\textcolor{red}{The geometrical details of the stable oxalate found in Ca-decorated graphene after re-optimization are C=C = 1.60 Å, C-O (coordinated to metal adatom) = symmetrical 1.32 Å, C-O (away from metal adatom) = 1.23 Å and the Ca-O distance = 2.10 Å}. 
 As shown in Fig. \ref{Binding_energy_plot}, the strongest adsorption energy is observed for the adsorption of 2 CO$_2$ molecules on Ca, Sr and Ti due to the oxalate formation in these systems. However, group 1 decorated graphene lacks the requisite reducing potential to convert CO$_2$ to a bent CO$_2$ anion. Consequently, oxalates do not form in these systems (see Fig. S3(c) and (d) in the SI), as group 1 metals can only donate one electron, whereas the formation of oxalate requires the donation of two electrons.

When 3-7 CO$_2$ molecules are adsorbed on the group 2 and Ti substrates, along with the oxalate formed, there are physisorbed CO$_2$ molecules in the system. %except for Ti@Gr where two oxalates formed when 4 CO$_2$ molecules are adsorbed (see Fig. S3(e) in SI).
The steady decrease in adsorption energy seen in Fig. \ref{Binding_energy_plot} is due to the averaging of the adsorption energy of the physisorbed CO$_2$ molecules and the oxalate formed in the systems.

To understand the adsorption strength of subsequent CO$_2$ on the M@Gr, we compute the step-wise adsorption energy for 1-5 CO$_2$ molecules on M@Gr by subtracting the total energy of physisorbed CO$_2$ on pristine graphene and the total energy of (n-1)CO$_2$ from the total adsorption energy of nCO$_2$ on M@Gr. %as given in Eqn.\ref{Eqn_2}. 
These energies are reported in the SI. The same trend seen for the average adsorption energy is seen for all the metal decorators. For group 1 decorated graphene, the adsorption energies are weak for 1-5 CO$_2$ molecules. On group 2 and Ti@Gr, 1-2 CO$_2$ molecules are strongly adsorbed, beyond this, the adsorption strength of subsequent CO$_2$ decreases to a physisorption energy regime.

%\begin{equation}\label{Eqn_2}
%     E^{stepwise}_{ads} = {E^{tot}_{\textit{n}CO_{2}//M@Gr} - E^{tot}_{\textit{(n-1)}CO_{2}//M@Gr} - E^{tot}_{CO_{2}//Gr}} 
%\end{equation}

%\noindent In Eqn. \ref{Eqn_2}, $E^{stepwise}_{ads}$ is the stepwise adsorption energy for CO$_2$ molecule, $E^{tot}_{\textit{n}CO_{2}//M@Gr}$ is the total energy of $n$CO$_2$ molecules adsorbed on M@Gr, $E^{tot}_{\textit{(n-1)}CO_{2}//M@Gr}$  is the total energy of the $(n-1)$CO$_2$ molecules adsorbed on M@Gr, $E^{tot}_{CO_{2}//Gr}$ is the total energy of physisorbed CO$_2$ molecule on pristine graphene and ${n}$ is the number of CO$_2$ molecules adsorbed. 

\subsection{Sticky oxalate}\label{subsec:sticky oxalate}
\begin{figure*}
    \centering  
    \includegraphics[width=1\textwidth]{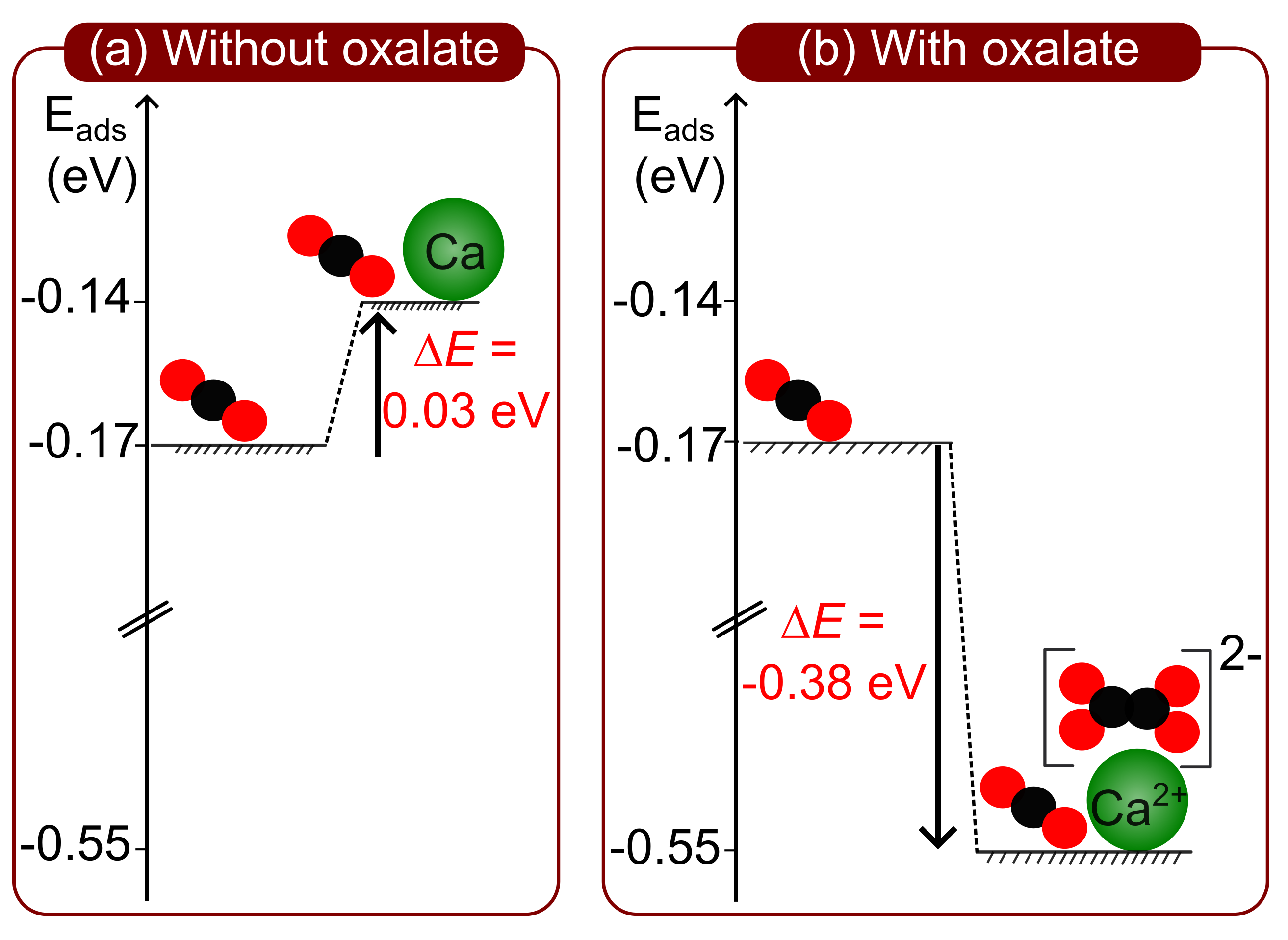}\captionsetup{justification=raggedright,singlelinecheck=false}
    \caption[ ]{Schematic representation of enhanced CO$_2$ adsorption in the presence of oxalate. (a) shows the energy difference from adsorbing a CO$_2$ on the pristine graphene to adsorbing it on a Ca-decorated graphene system, where the Ca and graphene have a $+$1e$^{-}$ and $-$1e$^{-}$  charge, respectively. (b) shows the energy change from adsorbing CO$_2$ on pristine graphene to adsorbing it on Ca@Gr in the presence of calcium-oxalate. It is worth noting that in systems where calcium oxalate forms it contributes -2.87 eV to the total adsorption energy of the system.} 
\label{sticky_oxalate_2}
\end{figure*}

In systems with more than 2 CO$_2$ molecules, we find that when oxalate is present, other CO$_2$ molecules bind more strongly. The role of the oxalate towards subsequently physisorbed CO$_2$ molecules is shown in Fig. \ref{sticky_oxalate_2}. On the left-hand side (Fig. \ref{sticky_oxalate_2}(a)), we show that bringing a physisorbed CO$_2$ towards a Ca adatom (with +1 valence charge) results in almost no change in energy. The $+$0.03 eV indicates that it is even more favourable to physisorb CO$_2$ on pristine graphene than on a monovalent Ca@Gr. It is worth noting that the monovalent Ca was from a single-point optimization.
However, the interaction of a physisorbed CO$_2$ with Ca-oxalate leads to an energy gain of about $-$0.38 eV as shown in Fig. \ref{sticky_oxalate_2}(b). This shows that the physisorbed CO$_2$ is more stabilized by the calcium-oxalate in the system compared to pristine graphene.

For the stepwise addition of the fourth CO$_2$  to oxalate on Ca@Gr, the CO$_2$ is stabilized by an energy of -0.33 eV. This further confirms that oxalate enhances the adsorption of CO$_2$. Upon adding the fifth CO$_2$ molecule, there is a $+$0.03 eV increase in energy, indicating that the adsorption of the fifth CO$_2$ molecule is destabilized, as it is located further away (about 8.95 Å) from the Ca-oxalate moiety in the system. Overall, these results reveal that for the third and fourth CO$_2$ molecules, the Ca-oxalate complex formed in this system is sticky, yielding a stronger interaction with physisorbed CO$_2$.

Our results show that group 2 decorated graphene and Ti-decorated graphene demonstrate stronger adsorption than group 1 metal decorators. This strength is due to the ability of metal atoms to reduce a linear CO$_2$ molecule to a bent CO$_2$ anion and form an oxalate upon the adsorption of 2 CO$_2$ molecules. The formation of the fully oxidized cation (+2) by the oxalate enhances the adsorption of multiple CO$_2$ molecules up to 4 CO$_2$ molecules for the group 2 decorated graphene system.

 %For practical CO$_2$ capture, the strong adsorption energy beyond the upper adsorption energy limit for ideal CO$_2$ capture seen for group 2 and Ti@Gr suggests these materials are not suitable for CO$_2$ capture at low concentrations where desorption is needed. However, owing to the relative abundance and low cost of group 2 metals, the material would be useful for permanent storage of CO$_2$. The strong adsorption energy will make the desorption of CO$_2$ and material regeneration difficult and prohibitively expensive. From a sustainability perspective, this makes the material a viable option for long-term CO$_2$ storage.
 
With our results, being aware of the limitation of the PBE-D3 functional used, we have performed additional calculations with other types of functionals: optB86b-vdW \cite{klimevs2011van}, optB88-vdW \cite{klimevs2009chemical}, SCAN+rVV10 \cite{peng2016versatile}, r2SCAN+rVV10 \cite{furness2020accurate}, as can be found in the SI. We find that the adsorption energies are shifted on the order of 0.25 eV, but the overall trends remain the same.  

\subsection{Chemisorption of \texorpdfstring {CO$_2$}{} as oxalate on decorated graphene}\label{sec:charge_transfer}

\begin{figure}[!ht]
    \centering \captionsetup{justification=raggedright,singlelinecheck=false}
    \includegraphics[width=1\textwidth]{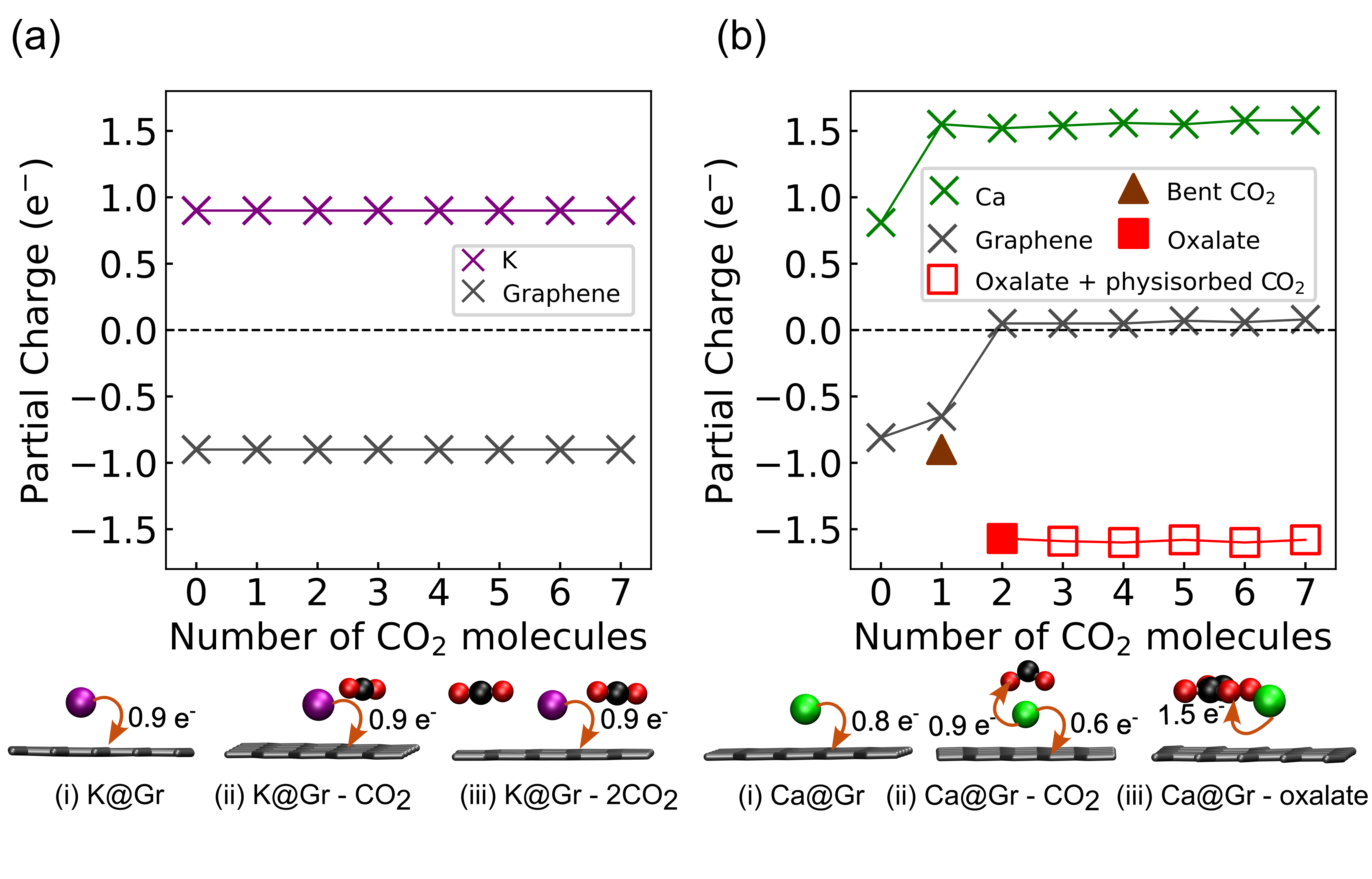}
    \caption[Charge analysis plot and schematic representation of charge transfer from metal adatom to CO$_2$ and graphene.]{Charge analysis plot and schematic representation of charge transfer from metal adatom to CO$_2$ and graphene. (a) shows the net charge transfer from K to the graphene sheet. The purple crosses show the charge on K and the grey crosses show the charge on the graphene sheet. (a)(i-iii) show the schematics of charge transfer for K@Gr, one CO$_2$ on K@Gr and 2 CO$_2$ molecules on K@Gr, respectively. In (b) the green and grey crosses show the charge on Ca, and graphene, respectively. The single brown triangle represents the charge transferred to the bent CO$_2$, the red-filled square represents the charge on the exclusive oxalate formed in the system and the unfilled red squares show the charge on the system where oxalate and physisorbed CO$_2$ are seen. (b)(i-iii) show the schematics of charge transfer for Ca@Gr, one CO$_2$ on Ca@Gr and 2 CO$_2$ molecules on Ca@Gr, respectively. For (a) CO$_2$ molecules are mainly physisorbed via van der Waals interactions with the  K@Gr surface, while in (b), the CO$_2$ molecules are chemisorbed strongly via ionic bonding with the Ca@Gr surface}
    \label{Bader charge analysis plot}
\end{figure}

 To gain insights into the redistribution of electrons and the nature of the bond formed between the CO$_2$ and the decorated graphene surfaces, a Bader charge analysis was performed on the Ca@Gr system as a representative system where an oxalate is formed and on K@Gr for group 1 decorated graphene system where no oxalate forms. Fig. \ref{Bader charge analysis plot} shows the change of net charge on graphene, K, Ca, and CO$_2$  molecules as the number of CO$_2$ increases from 0 to 7  CO$_2$ molecules. When K or Ca are adsorbed on the substrate, they reduce the graphene sheet by donating about one electron  (see Fig. \ref{Bader charge analysis plot} a(i) and b (i)).

 For 1-7 CO$_2$ adsorbed on K@Gr in Fig.~\ref{Bader charge analysis plot}(a), the 0.9 e$^{-}$ charge on the K adatom is donated to the graphene sheet. The fully oxidized  K adatom and the reduced graphene sheet are unable to reduce CO$_2$ to a bent anion. This explains the relatively weak adsorption energy on the group 1 decorated graphene compared to group 2 and Ti-decorated graphene. %However, the energy gained when one CO$_2$ interact with Ca$^{+1}$ as explained in case b of energy decomposition in section III is approximate to the adsorption energy of $-$0.41 eV seen for one CO$_2$ on K@Gr. This clearly confirms that although the reduction of CO$_2$ is not seen for this group of metal, a strong monovalent cation (K$^{+1}$)-CO$_2$ interaction relative to the pristine material is enhanced.

 %The CO$_2$ molecule is mainly physisorbed via van der Waals interactions in these systems.

Contrary to what is seen for K@Gr, when one CO$_2$ is adsorbed on Ca@Gr, CO$_2$ is reduced to a bent CO$_2$ anion by transfer of about one e$^{-}$ from Ca, and slightly less than one e$^{-}$  is donated from Ca to the graphene sheet, resulting in a positive charge of about 2 e$^{-}$ on Ca (see Fig. \ref{Bader charge analysis plot}b(ii)). For the adsorption of 2 CO$_2$ molecules where oxalate is formed, the roughly +2 e$^{-}$ on Ca is donated solely to the oxalate, showing an ionic bonding between the oxalate and the metal adatom. At 2-7 CO$_2$, where one oxalate is consistently formed, the roughly +2 e$^{-}$ on Ca is donated to oxalate entirely to form the Ca-oxalate complex and the graphene sheet becomes neutral (see Fig. \ref{Bader charge analysis plot} b(iii)).  %This is expected as the calcium oxalate is stabilized with a +1.5 charge from the Ca and a -1.5 charge from the oxalate in the gas phase for the BCA done in this study. 

In general, the direction of electron flow determines the adsorption strength of CO$_2$ on the substrate. When the valence electrons from the metal adatom are donated to the graphene, a relatively weak physisorption of CO$_2$ is seen. On the other hand, the complete loss of the valence electrons on the metal adatom to the adsorbed CO$_2$ molecule (as in the case of oxalate formation), results in a strong ionic chemisorption of CO$_2$  molecules on the substrate. These conclusions are consistent with a partial density of state analysis reported in the SI.
%To further analyse the electronic structure of the CO$_2$ capture on Ca-decorated graphene surface, the partial density of the state of 1-2 CO$_2$ molecule on Ca-decorated graphene was examined. The O-p state of the adsorbed CO$_2$ molecule(s) is populated as a result of charge transfer from the metal adatom. However, no hybridization of the electronic p-state and the s,d metallic states of the Ca adatom near the fermi energy was seen in our system, as reported in Ref. \cite{Cazorla2011}.
 
\section{Discussion}\label{sec:discussion}
%Low-energy structures are the most important for predicting material properties because they are more abundant under equilibrium conditions and 
%more easily accessible experimentally. However, identifying them in a complex system is challenging for experimentalists \cite{pickard2011ab, oganov2019structure, mcmahon2011ground}. 
 %With the RSS approach, a large chemical space can be effectively searched to obtain interesting structures and discover promising new and stable materials 

Through comprehensive exploration of the potential energy surface with RSS, we found an oxalate as the lowest energy structure in group 2 and Ti@Gr. To the best of our knowledge, adsorption structures such as the oxalate have not been reported in the few studies that investigated multiple CO$_2$ molecule adsorption on metal-decorated or doped graphene surfaces \cite{tawfik2015multiple}.

Previous theoretical studies have reported enhanced adsorption due to the reduction of one CO$_2$ molecule to a bent CO$_2$ anion on Ca-decorated surfaces. For instance, Tawfik \textit{et al.} \cite{tawfik2015multiple}, found only the first adsorbed CO$_2$ to bend on single and double vacancy  Ca@Gr with strong adsorption energies of $-$1.10 eV and $-$1.30 eV, respectively. Upon the subsequent addition of CO$_2$, there was no further reduction of the CO$_2$ molecules \cite{tawfik2015multiple}. However, in our work, 2 CO$_2$ molecules are reduced to oxalate on group 2 decorated graphene systems and two oxalates are formed when 4 CO$_2$ molecules are adsorbed on Ti-decorated graphene without the presence of defects.

There are many possible reasons why an oxalate was not formed in Ref. \cite{tawfik2015multiple}.  First, differences in the system, such as unit cell size and defects on the graphene \cite{tawfik2015multiple}. More importantly, previous work might have missed the oxalate formation because wider configurations space of CO$_2$ on metal-doped graphene were not sampled with sampling techniques such as the RSS. The latter is supported by the observation that the CaO clusters reported in Ref. \cite{tawfik2015multiple} were seen in some of the high-energy structures discovered by the RSS technique performed in our work.

Experimental works \cite{angamuthu2010electrocatalytic, marx2022revisiting, rudolph2000macrocyclic} have reported the electrocatalytic conversion of CO$_2$ to oxalate by transition metals.
%Angamuthu\textit{ et al., }\cite{angamuthu2010electrocatalytic} have found a copper complex to demonstrate remarkable selectivity in the reductive coupling of CO$_2$ to form oxalate through coordinative electron transfer. 
The reduction of CO$_2$ to bent CO$_2$ anion is the first step to the conversion of CO$_2$ to higher C$_2$ compounds such as oxalate, however, the large reorganisation energy needed for the reduction of CO$_2$ to bent CO$_2$ anion makes it challenging for experiments. In this work, our results demonstrate that the RSS technique could be a suitable modelling tool %for experimentalists 
to screen for bent CO$_2$ molecules and other higher C$_2$ compounds like oxalate when investigating CO$_2$ reduction on low-dimensional materials.

\section{Conclusion and outlook}\label{sec:conclusion}

By systematically exploring a relatively wide configuration space of CO$_2$ molecules on a decorated graphene system with random structure search, we have gained the following insights: (i) There is an enhanced adsorption of CO$_2$ molecules on M@Gr compared to pristine graphene; (ii) Oxalates form on group 2 and Ti-decorated graphene. Oxalate formation in these systems gave the strongest adsorption energy of CO$_2$ on group 2 and Ti-decorated graphene surface and facilitated the cooperative adsorption of other physisorbed CO$_2$ molecules; (iii) The adsorption mechanism observed for the enhanced CO$_2$ adsorption is based on charge transfer, and the charge transfer determines the adsorption strength. Strong chemisorption is seen when charges from metal atoms are transferred to the CO$_2$ molecules, and relatively weak physisorption is seen when all the electrons on the metal atom are transferred to graphene; (iv) Oxalate can enhance the adsorption of additional CO$_2$ molecules to the metal-decorated graphene surface.

This study is primarily aimed at understanding how CO$_2$ interacts with decorated graphene surfaces, and many factors that might affect the practical usage of these materials for CO$_2$ capture have not been considered because they are beyond the scope of this current work. However, the fundamental insights about the formation of oxalate and its cooperative effect in enhancing the adsorption of other CO$_2$ molecules suggest a potential strategy for higher CO$_2$ uptake on decorated graphene systems. It would be extremely valuable to confirm the formation of oxalates experimentally while considering other realistic factors such as the effect of defects, water, and temperature on these types of materials.

\section*{Acknowledgments}
    I.C. acknowledges the support from the Bill \& Melinda Gates Foundation [OPP1144]. Y.S.A. is supported by Leverhulme grant no. RPG-2020-038. F.B. acknowledges support from the Alexander von Humboldt Foundation through a Feodor Lynen Research Fellowship, from the Isaac Newton Trust through an Early Career Fellowship, and from Churchill College, Cambridge, through a Postdoctoral By-Fellowship. B.X.S. acknowledges support from the EPSRC Doctoral Training Partnership (EP/T517847/1). A.Z. and D.A. acknowledge support from the European Union under the Next generation EU (projects 20222FXZ33 and P2022MC742).
 Calculations were performed using the Cambridge Service for Data-Driven Discovery (CSD3) operated by the University of Cambridge Research Computing Service (www.csd3.cam.ac.uk), provided by Dell EMC and Intel using Tier-2 funding from the Engineering and Physical Sciences Research Council (capital grant EP/T022159/1 and EP/P020259/1). We additionally acknowledge computational support and resources from the UK national high-performance computing service, Advanced Research Computing High-End Resource (ARCHER2). Access for ARCHER2 was obtained via the Materials Chemistry Consortium (MCC), funded by EPSRC grant reference EP/X035859.

\section*{REFERENCES}    
%\bibliography{apssamp}

\bibliography{references}

\newpage
\setcounter{section}{0}
\renewcommand{\thesection}{S\arabic{section}}%
\setcounter{table}{0}
\renewcommand{\thetable}{S\arabic{table}}%
\setcounter{figure}{0}
\renewcommand{\thefigure}{S\arabic{figure}}%
\section*{Supporting Information}

\section*{Summary}

In this supporting information, we provide: (i) The total adsorption energy for 1-7 CO$_2$ adsorbed on metal decorated graphene surfaces, and the optimised geometries for CO$_2$ on pristine graphene and 1-7 CO$_2$ on all M@Gr investigated; (ii) Give the stepwise adsorption energies for 1-5 CO$_2$ molecules on all M@Gr studied; (iii) Provide the density of state analysis for 1 CO$_2$ on K@Gr and 1-2 CO$_2$ molecules on Ca@Gr; (iv) Give the convergence test details for the energy cut-off and k-point mesh used in the study;  (v) Show the insensitivity of the adsorption trends in our work to functional choice; and (vi) Provide the sample of VASP input file for the optimization of 2 CO$_2$ on Ca@Gr along with coordinates of all the optimised geometries of 1-7 CO$_2$ molecules on M@Gr that might be needed to reproduce this work.

\section{Total adsorption energy of \texorpdfstring{CO$_2$}{} on decorated graphene systems}
\begin{figure}[!ht]
    \centering
    \includegraphics[width=0.5\textwidth]{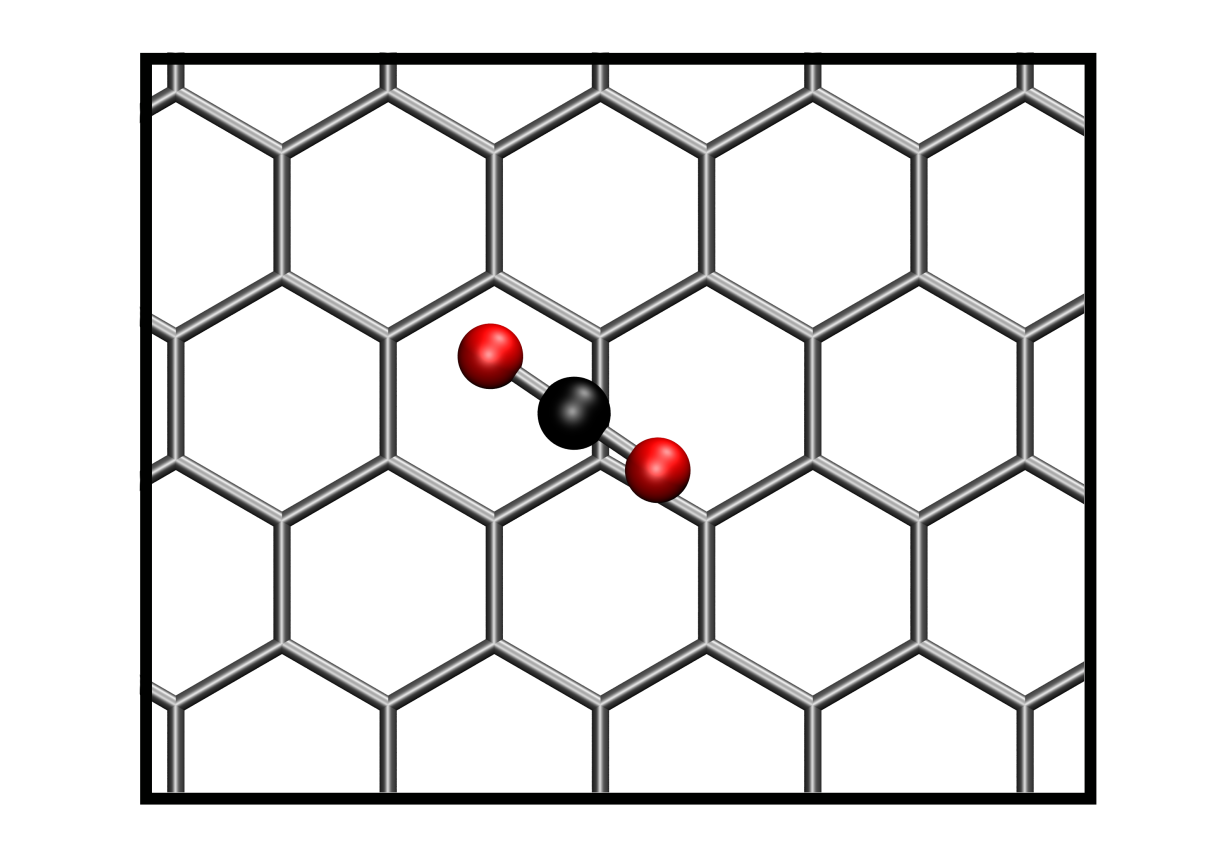}
    \caption[Optimised structures of CO$_2$ on pristine graphene and 2CO$_2$ on K@Gr]{Optimised orientation of CO$_2$ on pristine graphene found with our RSS approach.}
    \label{graphene_inset}
\end{figure}

CO$_2$ has been previously reported to physisorb on pristine graphene just as we also found in this work. The optimised geometry of CO$_2$ on pristine graphene from our random structure (RSS) result is shown in Fig. \ref{graphene_inset}.

\begin{figure}%[!htbp]
    \centering
    \includegraphics[width=0.5\textwidth]{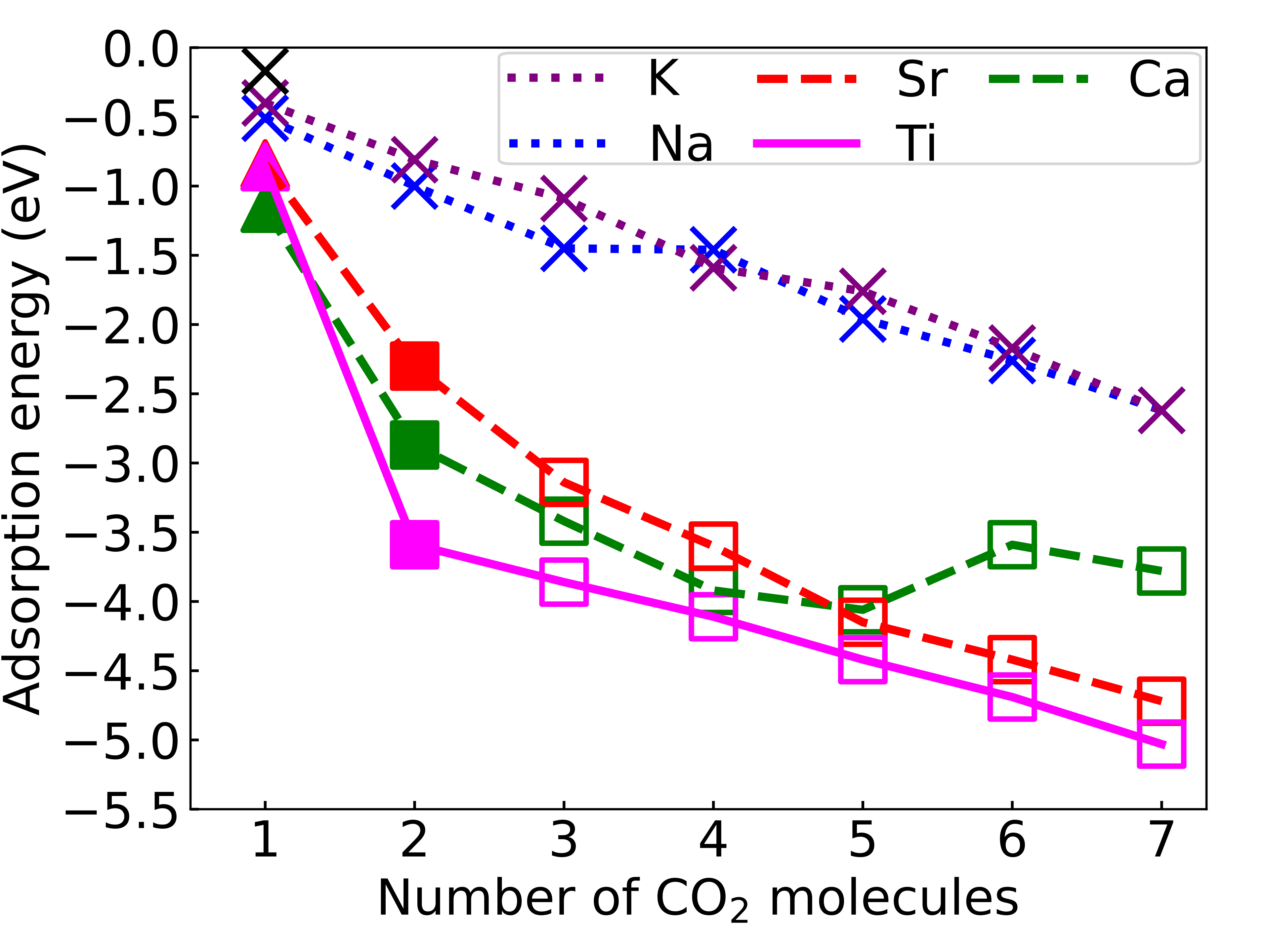}
    \caption[Total adsorption energy plot of 1-7 CO$_2$ molecules adsorbed on Ca, Sr, Na, K and Ti-decorated graphene]{Total adsorption energy of 1-7 CO$_2$ molecules adsorbed on Ca, Sr, Na, K and Ti@Gr. The symbol indicates the optimized orientation of CO$_2$ molecules around the metal atoms. The triangles indicate the formation of a bent CO$_2$ anion, the filled squares indicate the exclusive formation of oxalate, the unfilled squares represent the system where there is an oxalate and physisorbed CO$_2$ molecules, the crosses indicate the radial orientation of the physisorbed CO$_2$ molecules to the metal atom. The dotted lines represent group 1 elements; the dashed lines indicate group 2 elements and the solid line shows the transition metal Ti.}
    \label{Total_Binding_energy_plot}
\end{figure}

\begin{figure}%[!htbp]
    \centering
    \includegraphics[width=1\textwidth]{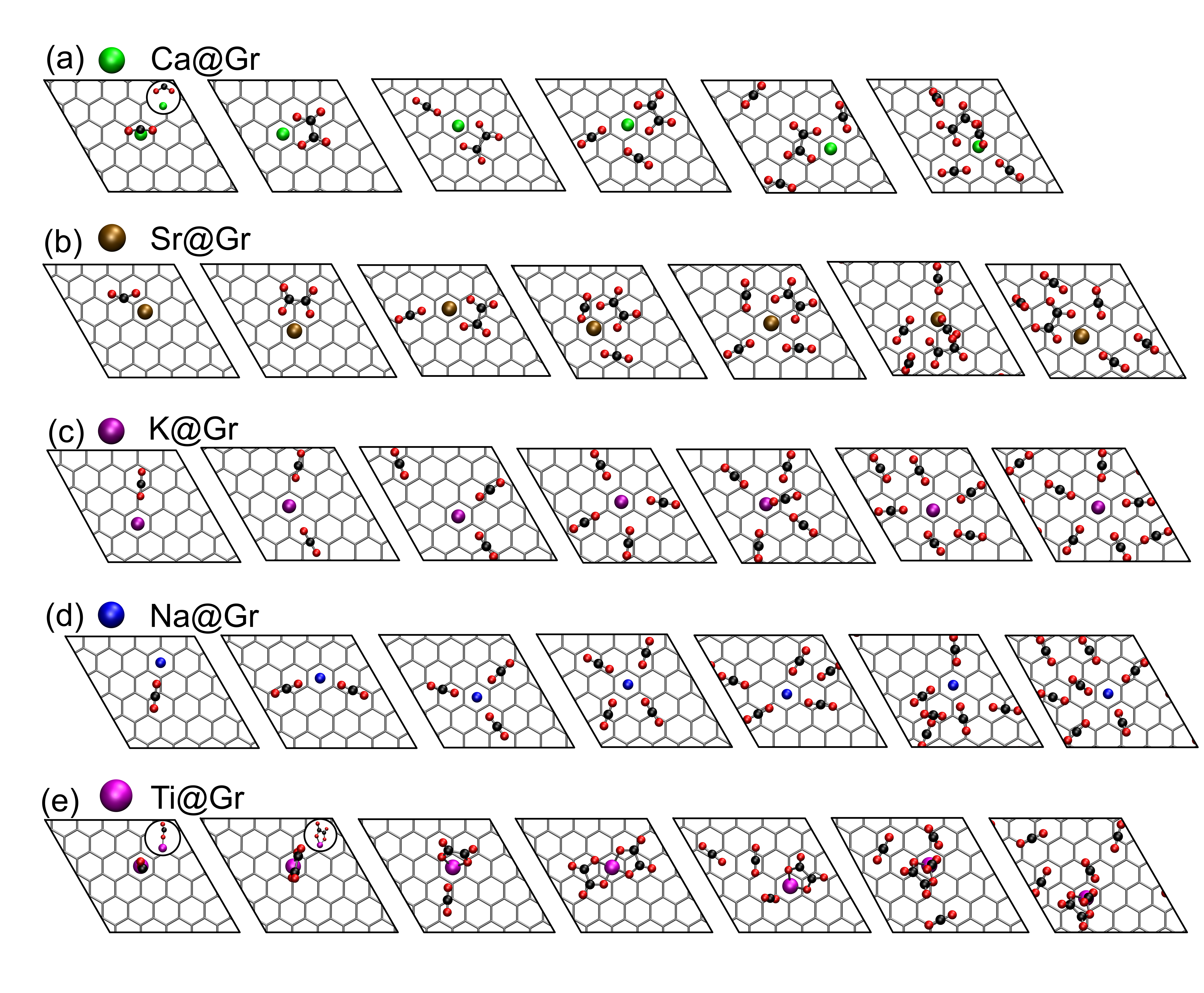}
    \caption[ ]{Optimized geometries of 1 -7 CO$_2$ on (a) Ca, (b) Sr, (c) K, (d) Na, and (e) Ti decorated graphene systems. The first adsorbed CO$_2$ is seen to bend on Ca, Sr and Ti@Gr. From the adsorption of 2-7 CO$_2$ molecules, an oxalate is formed alongside with other physisorbed CO$_2$ molecules. For 4 CO$_2$ molecules on Ti@Gr, 2 oxalates are stabilized on the system after which just one oxalate remains as CO$_2$ molecules increases from 5-7. For all number of CO$_2$ molecules adsorbed on Na and K@Gr, CO$_2$ molecules are radially coordinated to the metal adatoms and no bent CO$_2$ or oxalate is formed at higher loading of CO$_2$ molecules on these systems.}
    \label{optimised structures}
\end{figure}

In the main manuscript, we reported the average adsorption energies for 1-7 CO$_2$ molecules on the metal-decorated systems. The average adsorption energy decreases as the number of CO$_2$ molecules increases from 3-7 CO$_2$ molecules because there are more weakly bounded CO$_2$ in the system. Here, we show the total adsorption energy without normalization, for 1-7 CO$_2$ molecules adsorbed on the metal-decorated graphene. As seen in Fig. \ref{Total_Binding_energy_plot},
 the total adsorption energy increases for all the systems investigated, with Ti demonstrating the strongest adsorption, followed by Ca, Sr, Na and K. The optimised geometries for 1-7 CO$_2$ molecules are shown in Fig. \ref{optimised structures}.

\section{Step-wise adsorption energies of \texorpdfstring{CO$_2$}{} on decorated systems}

\begin{figure}%[!ht]
    \centering
    \includegraphics[width=0.5\textwidth]{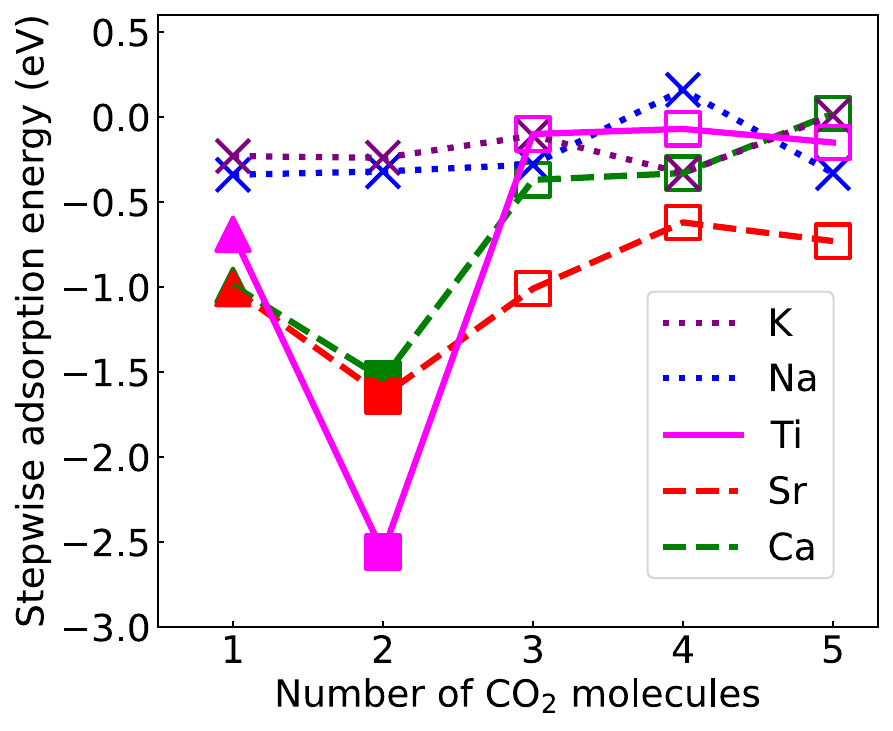}
    \caption[Stepwise adsorption energy of 1-5 CO$_2$ molecules adsorbed on Ca, Sr, Na, K and Ti-decorated graphene]{Stepwise adsorption energy of 1-5 CO$_2$ molecules adsorbed on Ca, Sr, Na, K and Ti@Gr. The symbol indicates the optimized orientation of CO$_2$ molecules around the metal atoms. The triangles indicate the formation of a bent CO$_2$ anion, the filled squares indicate the exclusive formation of oxalate, the unfilled squares represent the system where there is an oxalate and physisorbed CO$_2$ molecules, the crosses indicate the radial orientation of the physisorbed CO$_2$ molecules to the metal atom. The dotted lines represent group 1 elements; the dashed lines indicate group 2 elements and the solid line shows the transition metal Ti.}
    \label{stepwise_adsorption_energy_plot}
\end{figure}

The trends seen for the average adsorption energies of 1-7 CO$_2$ on M@Gr reported in the main manuscript is found to hold when we compute the step-wise adsorption energies for 1-5 CO$_2$ on M@Gr systems (see Fig. \ref{stepwise_adsorption_energy_plot}). CO$_2$ molecule is most stable when oxalate is formed on Ca, Sr and Ti@Gr, and the stability of subsequent physisorbed CO$_2$ molecules is enhanced compared to CO$_2$ molecule adsorbed on pristine graphene. This stability decreases at higher CO$_2$ loading as more CO$_2$ are located further away from the metal adatom- oxalate complex. For K and Na@Gr systems, the step-wise adsorption energies remain in the physisorption regime.

\section{Projected Density of states}

\begin{figure}%[!htbp]]
    \centering
    \includegraphics[width=0.5\textwidth]{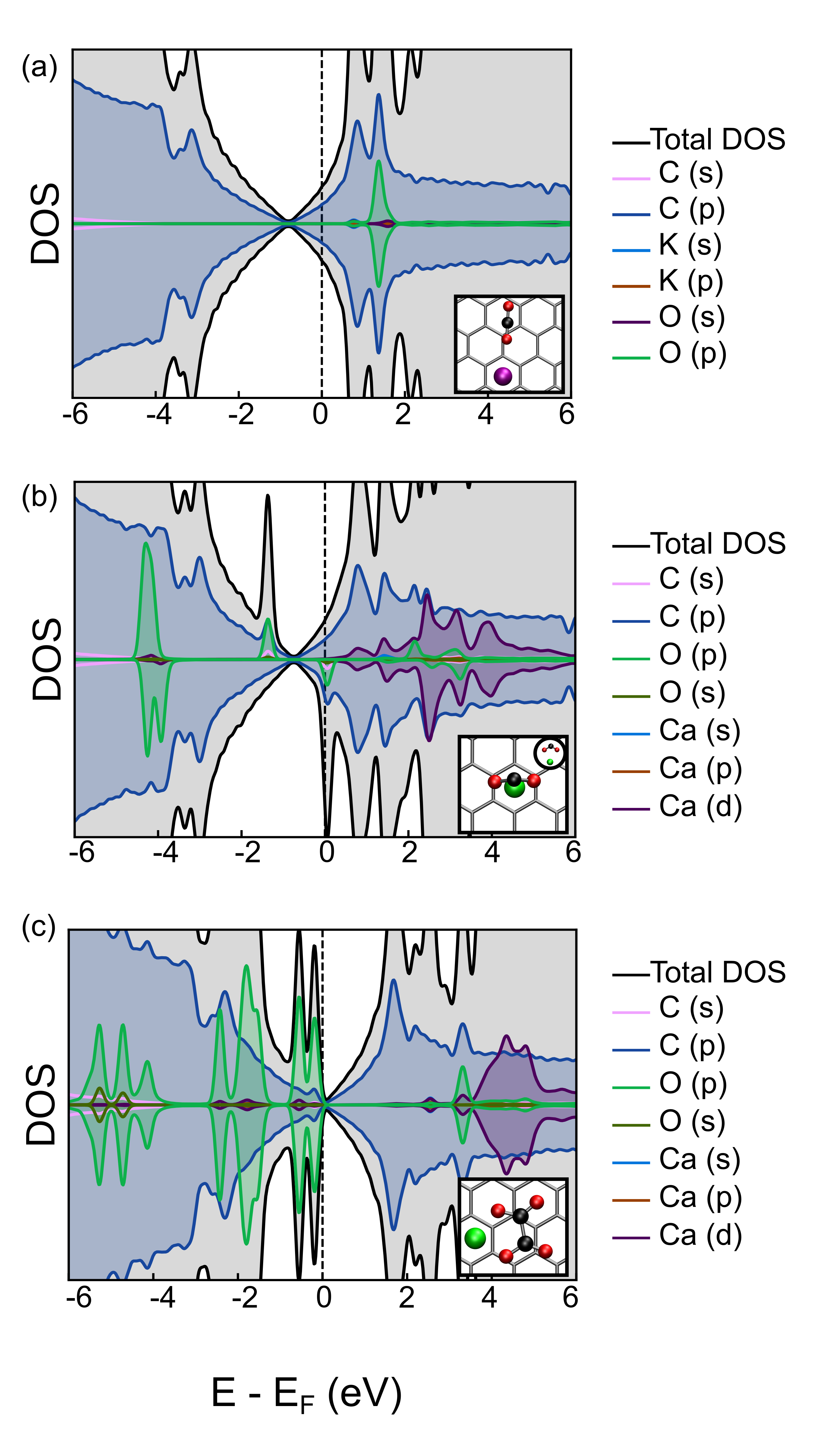}
    \caption{The projected density of states (PDOS) within $\pm$ 6 eV of the Fermi energy for 1 CO$_2$ on K@Gr, 1 CO$_2$ on Ca@Gr and 2 CO$_2$ on Ca@Gr. The PDOS has been shifted to the Fermi level for each systems. The grey shaded region indicates the total DOS and the corresponding colour of state projections, as shown in each system's label. A schematic of the configuration of CO$_2$ on the metal atom is shown in the insets.  }
    \label{dos plot}
\end{figure}

 Graphene is described as semi-metal because of its vanishing density of state at the Fermi level \cite{malko2012competition,ferrari2006raman}. This unique property makes its electronic properties easily tunable with few charge transfers. Introducing different metal adatoms changes the electronic properties of graphene as CO$_2$ is adsorbed (see Fig \ref{dos plot}). When 1 CO$_2$ is adsorbed on K@Gr and Ca@Gr, the transfer of about 1 e$^{-}$ from K and Ca to Gr, shifted the DOS from the Fermi level to lower energy where it is slightly more metallic than its pristine form as seen Fig \ref{dos plot} (a) and Fig \ref{dos plot} (b). The population of the O-p state at about -1 eV (in Fig. \ref{dos plot} (b)) after the formation of the bent CO$_2$ anion on Ca@Gr indicates that O has been reduced following the formation of the bent CO$_2$ anion as charges are transferred from Ca to the adsorbed CO$_2$. This suggests the underlying mechanism here as an ionic charge transfer from the metal atom to the CO$_2$ molecules.

In  Fig. \ref{dos plot} (c), where oxalate is formed on Ca@Gr, it could be seen that graphene maintains its semi-metallic nature by having a vanishing density of state at the Fermi energy. The population of O-p state below the Fermi level shows that oxygen has been reduced as it takes charge from Ca, to form the Ca-oxalate complex. This suggests that the mechanism here also was a charge transfer that led to ionic bonding between the cation of the metal atom and the oxalate anion formed. This conclusion is consistent with the Bader charge transfer reported for these systems in the main manuscript.

\section{Convergence tests}

\begin{figure}%[!ht]
    \centering
    \includegraphics[width=0.5\textwidth]{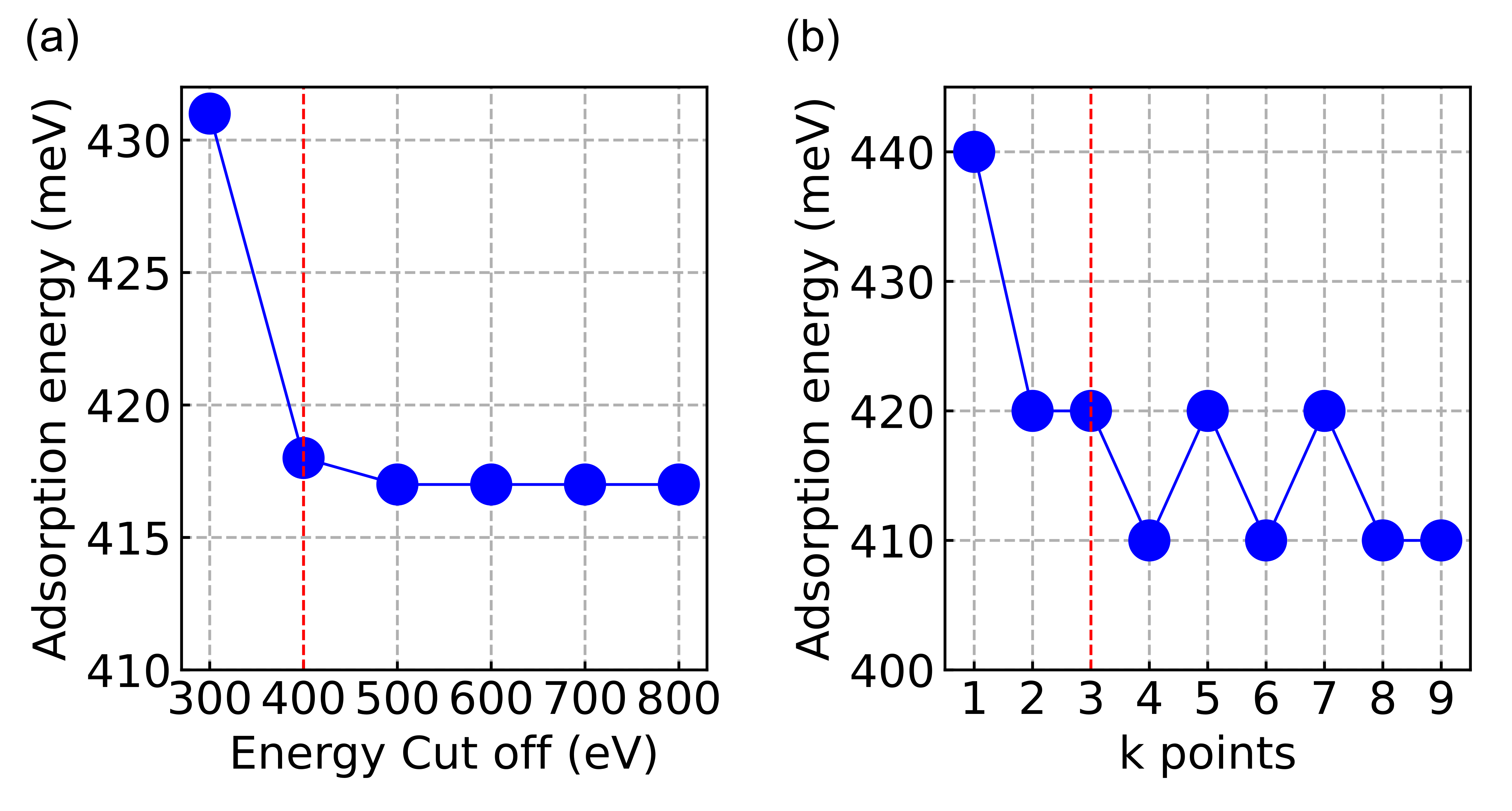}
    \caption{Convergence test for (a) the energy cut-off and (b) \textbf{k}-point mesh used in this study. The x-axis of Fig. \ref{covergence_test} b is not the number of \textbf{k}-points but the value of 1$\times$1$\times$1 to 9$\times$9$\times$1 \textbf{k}-point mesh. The red vertical line in a and b is the \textbf{k}-point mesh and energy cut-off used in our calculations.}
    \label{covergence_test}
\end{figure}

To know the effective point to truncate the infinite plane waves used in these calculations, and the efficient \textbf{k}-point mesh to sample the reciprocal space, a convergence test was done for energy cut-off values of (300 - 800 eV) and \textbf{ k}-point mesh of (1$\times$1$\times$1 to 9$\times$9$\times$1) for a representative system containing one CO$_2$ molecule on calcium decorated graphene (Ca@Gr). The energy cut-off of 400 eV is found to converge within 1meV of the 800 cut-off, and  3x3x1 k point converged within 10 meV of the 9$\times$9$\times$1 k-point mesh (see Fig. \ref{covergence_test}). Therefore,
the energy cut-off of 400 eV and 3$\times$3$\times$1 \textbf{k}-points mesh are used for the geometry relaxation of the structures in this study.

\section{Sensitivity of oxalate formed on M@Gr to functional choice}

To establish the reliability of the adsorption trends seen with the PBE-D3 functional used in this work, particularly, the formation of oxalate when 2 CO$_2$ is adsorbed on group 2 and transition metal decorated graphene system, we computed the adsorption energy of 1-3 CO$_2$ on Ca@Gr as a representative system where oxalate is formed using four other functionals namely; optB86b-vdW \cite{klimevs2011van}, optB88-vdW \cite{klimevs2009chemical}, SCAN+rVV10 \cite{peng2016versatile}, r2SCAN+rVV10 \cite{furness2020accurate}. Since London dispersion interaction could contribute to the stability of molecules adsorbed on surfaces, we chose these functionals because they account for nonlocal effects and give more accurate interaction energies for the description of adsorbate-substrate interactions \cite{klimevs2011van}.  The adsorption energies of these functionals are reported in Table \ref{Functional_tests}. Even though the meta-GGA adsorption energies are more stable for the systems investigated, oxalate is still found with these functionals. This shows that PBE-D3 is sufficient in describing the interaction of CO$_2$ molecules on metal-decorated graphene surfaces which gives confidence in our results.

\begin{table}%[!ht]
\centering
\caption{Comparison of the PBE-D3 average adsorption energy of 1-3 CO$_2$ on Ca@Gr with other functionals.}
\label{Functional_tests}
\begin{tabular}{@{}ccc@{}}
\toprule
Functional & Systems & E$_{ads}$(eV) \\ \midrule
PBE+D3 & 1CO$_{2}$ & -1.16 \\
 & 2CO$_{2}$ & -1.44 \\
 & 3CO$_{2}$ & -1.14 \\
 &  &  \\
optB86b-vdW & 1CO$_{2}$ & -1.17 \\
 & 2CO$_{2}$ & -1.60 \\
 & 3CO$_{2}$ & -1.27 \\
 &  &  \\
optB88-vdW & 1CO$_{2}$ & -1.20 \\
 & 2CO$_{2}$ & -1.61 \\
 & 3CO$_{2}$ & -1.29 \\
 &  &  \\
SCAN+rVV10 & 1CO$_{2}$ & -1.40 \\
 & 2CO$_{2}$ & -1.67 \\
 & 3CO$_{2}$ & -1.32 \\
 &  &  \\
r$^{2}$SCAN+rVV10 & 1CO$_{2}$ & -1.34 \\
 & 2CO$_{2}$ & -1.60 \\
 & 3CO$_{2}$ & -1.27 \\ \bottomrule
\end{tabular}
\end{table}

\section{INPUT FILES FOR VASP}
Some example input files are given here to guide the interested reader. Example of VASP input files for optimising the structure of 2 CO$_2$ molecules on Ca@Gr:

ENCUT = 400

EDIFF = 1E-6

ALGO = Fast

PREC = Accurate

NELM = 125  

ISMEAR = 0   

SIGMA = 0.05  

ISPIN = 2

MAGMOM = 52*0 1*2 4*0

IBRION = 2  

NSW = 500  

ISIF = 0   

EDIFFG = -1E-2

NWRITE = 2

LWAVE = .FALSE.

GGA    = PE   

IVDW = 11

%\section*{REFERENCES}
%\bibliographystyle{ieeetr}

%\bibliography{07_Seventh_draft_arxiv/SI/si_references}

\end{document}